\documentstyle[psfig]{mn}

\title[Fundamental galaxy parameters for radio-loud AGN]{Fundamental galaxy parameters for radio-loud AGN and the black hole $-$ radio power connection}

\author[I. Snellen et al.]
{I.A.G. Snellen$^{1}$, M.D. Lehnert$^2$, M.N. Bremer$^3$,
R.T. Schilizzi$^{4,5}$\\ 
$^{1}$ Institute for Astronomy, Blackford Hill, Edinburgh EH9 3HJ, United Kingdom\\
$^{2}$ Max-Planck-Institut f\"ur extraterrestrische Physik (MPE), Postfach 1312, 85741 Garching, Germany\\
$^{3}$ Department of Physics, Bristol University, H H Wills Physics Laboratory, Tyndall Avenue, Bristol, BS8 1TL, United Kingdom\\
$^{4}$Joint Institute for VLBI in Europe, Postbus 2, 7990 AA, Dwingeloo, 
The Netherlands\\
$^{5}$Leiden Observatory, P.O. Box 9513, 2300 RA, Leiden, The Netherlands \\
}

\date{}

\begin{document}

\maketitle

\begin{abstract}
We have determined the central velocity dispersions and surface 
brightness profiles for a sample of powerful radio galaxies 
in the redshift range $0.06<z<0.31$, which were selected on the basis of 
their young radio source.
The optical hosts follow the fundamental plane of elliptical galaxies, 
showing that young radio sources reside in normal ellipticals, 
as do other types of radio galaxies into which these objects are believed to 
evolve. As young radio sources are relatively 
straightforward to select and the contributions of the AGN light
to the optical spectra are minimal, these objects
can readily be used to study the evolution of the fundamental plane
of elliptical galaxies out to z=1, independently of optical selection effects.

The black hole masses, $M_{bh}$, of the objects in our sample have been
determined using the tight empirical relation of $M_{bh}$ with central
velocity dispersion, $\sigma_e$, and for literature samples of classical radio 
galaxies and optically selected ellipticals. Only the optically selected 
in-active galaxies are found to exhibit a correlation between $M_{bh}$ and 
radio luminosity. In contrast, the radio powers of the AGN in 
the samples do not correlate with $M_{bh}$ at all, with objects at 
a given black hole mass ranging over 7 orders of magnitude in 
radio power.

We have been able to tie in the local population of powerful radio sources
with its parent population of in-active elliptical galaxies:
the local black hole mass function has been determined using 
the elliptical galaxy luminosity function and the Faber-Jackson 
and $M_{bh}-\sigma$ relations. This was combined with
the fraction of radio-loud black holes as function
of $M_{bh}$, as determined from the optically selected galaxy sample,
to derive the local volume-density of radio galaxies and the
distribution of their black hole masses.
These are shown to be consistent with the local radio
luminosity function and the distribution of black hole masses as found in 
the radio selected samples, and confirms that elliptical 
galaxies comprise the large majority of the radio-loud population of active galaxies.
\end{abstract}


\section{Introduction}
\subsection{Galaxy formation and the fundamental plane}
How and when galaxies were formed and how they have
evolved since, is one of the central issues of contemporary
astrophysics today. 
Elliptical galaxies may be the best objects to study galaxy evolution,
since
their optical light is not dominated by relatively recent
star-formation, unlike that of spiral galaxies, and their
observational properties are a better reflection of their
long-timescale cosmological development.

In the 1980s it was shown that the main global
properties of local elliptical galaxies, velocity dispersion, surface
brightness and effective radius, are strongly correlated, and that 
they describe a plane in this 3-dimensional parameter space
(Dressler et al., 1987; Djorgovski \& Davis, 1987). 
Initially, this `fundamental plane' (FP) of elliptical galaxies drew
attention mainly for its use as a distant indicator, to measure the
Hubble constant and peculiar motions in the local universe. However,
it was also realised that it implied a strong regularity
in the process of galaxy formation, in particular that the 
mass-to-light (M/L) ratio varied only slightly over 4 orders of
magnitude in galactic mass (Djorgovski \& Davis, 1987).

Since the M/L ratio is a strong function of the age of the stellar population, 
the study of the fundamental plane as function of redshift is a
sensitive diagnostic of the evolutionary history of galaxies. 
van Dokkum, Franx and Kelson and collaborators, 
who observed elliptical cluster galaxies with 
the Keck telescope and the HST, located at z=0.39 (van Dokkum \& Franx 1996),
z=0.58 (Kelson et al. 1997), and z=0.83 (van Dokkum et al. 1998),
showed that the M/L ratio evolves as $\Delta$ log $M/L_B \propto
-0.4z$ ($\Omega_m=0.3$,$\Omega_\Lambda = 0$), putting strong 
constraints on the formation epoch of cluster ellipticals, with
$z_{form}>2.8$ for a Salpeter initial mass function.

The Mg\,b or Mg$_2$ indices are also valuable measures of the age
of the stellar population, which is often used in conjunction with the
FP.  The Mg$_2$ index is found to be correlated with the central 
velocity dispersion in local elliptical galaxies, 
with only a small scatter (eg. Dressler et al., 1987; Djorgovski \&
Davis, 1987, J\/orgensen, Franx \& Kjaergaard 1996). 
Ziegler \& Bender (1997) showed that for 
cluster galaxies at z=0.37, this correlation is offset by $<$Mg\,b$>$
$\approx -0.4$\AA\, which can be fully attributed to the lower 
luminosity-weighted ages
of these stellar populations at this earlier epoch.

These studies may seem to suggest that, except for perhaps some
minor details, the evolution of elliptical galaxies is well
understood. However,
various strong biases may still be present in the studies described above. 
The most serious bias may
be the progenitor bias (van Dokkum \& Franx 1996): present day
ellipticals which were spirals at an earlier epoch, would not have been 
included in a sample at that epoch, biasing the sample to what are 
likely to be the oldest galaxies. Furthermore, there is also an
environmental bias: present studies are focused on 
cluster galaxies, while it is not known how these clusters evolve with
cosmological epoch. For example, it is far from clear what a local
cluster such as Coma would have looked like at high redshift, and to
what present-day structures the high-z clusters have evolved into.
More generally, connecting populations at one redshift
with their predecessor at high redshift, without making uncertain
assumptions about the history of star formation is a problem.

\subsection{The role of nuclear activity in galaxy formation and evolution}

The place of active galaxies in the general picture of 
galaxy evolution is unclear. Until recently, 
AGN were seen as rare and
peculiar objects, which, although interesting for their own sake, 
were
disconnected from galaxy evolution issues. However, the recent 
discovery that every galaxy seems to have a central massive black 
hole, the mass of which is closely related to the total
mass of (the bulges of) its host (Magorrian et al, 1998;
Gebhardt et al. 2000), has changed this perception dramatically. 
It is now clear that the formation and evolution of the 
central massive black hole is closely linked to that of the 
host galaxy (eg. Silk \& Rees 1998).
In addition, it implies that all galaxies may be capable of having 
phases of central activity, meaning that active galaxies are just 
normal galaxies, with their central black hole coincidently caught
during a period of (high) accretion. This places active galaxies 
 back in the centre of attention of galaxy formation and 
evolution studies, especially since they are relatively easy to select 
at  high redshift. Recent claims suggest that non-thermal radio emission
from ellipticals is strongly correlated with the mass of their central 
black hole, ranging from dwarf ellipticals to powerful radio galaxies 
(Franceschini et al. 1998). This would make the connection between quiescent and active galaxies even stronger. 

\subsection{Radio galaxies as probes of galaxy evolution}

Since active galaxies play such a potentially important role in
galaxy evolution, it is important to determine their 
fundamental structural parameters as function of redshift.
An additional factor is that selection of AGN allows the study of galaxy
evolution independently of the optical selection biases as mentioned above. 
Smith, Heckman \& Illingworth (1990) have studied the fundamental
plane parameters of nearby powerful radio galaxies, and showed 
that they indeed reside in normal ellipticals. A similar 
result has been found recently by Bettoni et al. (2002). 
Unfortunately, it is almost impossible to use similar
objects at higher redshift, since their 
spectral energy distributions have invariably been found to be
strongly influenced by AGN light, especially in the optical
(eg. optical/UV alignment effect, McCarthy et al. 1987, Chambers et
al. 1987). 

This problem can be avoided by selecting galaxies containing a very 
young radio source. In the very early stages of radio source 
evolution, when the radio source is only 100-1000 years old and 
confined to the central hundred parsecs, the 
contribution of the AGN to the global properties of its host galaxy 
is still minimal (Snellen et al. 1996a,b; 1999): broadband photometry showed 
that their optical to 
near-infrared colours are consistent with passively evolving
ellipticals, indicating that their light is dominated by old stellar
populations. The dispersion in their $R$ and $K$ band
Hubble diagrams (0.3 mags) is smaller than that for other 
classes of radio galaxies, indicating a homogeneous and well behaved
population of hosts (Snellen et al. 1996a,b). In addition, their optical 
spectra typically only show weak emission lines but deep stellar
absorption lines, again indicative of them being dominated by old stellar
populations (Snellen et al. 1999). For example, the archetype young
radio source, B0108+388, is as powerful as Cygnus A. However,
it has a remarkably low OIII$_{5007}$ equivalent
width of only $\sim$4\AA$ $(Lawrence et al. 1996), corresponding
to a line luminosity about a factor 50 lower than that of Cygnus A.

That the contamination from young radio-loud AGN is only small or 
even absent may not come as a surprise. The young radio source may
not yet have pierced through the dense and dusty central regions of
the host. AGN related light is also likely to still be confined to 
the inner hundreds of light years, and obscured from our view.
There is indeed much evidence through HI absorption studies of young 
radio-loud AGN that large column densities of gas are present 
(eg. Carilli et al. 1999; Peck, Taylor \& Conway 1999)

In this paper, we investigate the use of the host galaxies of 
young radio-loud AGN as probes of galaxy evolution. The 
connection between radio activity and black hole mass, as infered
from their central velocity dispersion, has also been investigated. 
We present high dispersion
spectroscopic and optical CCD imaging data of a sample of 7 young
radio-loud AGN located between $0.06<z<0.31$. Section 2 describes
 the properties of young radio-loud AGN and the selection
of the sample. Section 3 describes the observations, reduction and 
analysis. Section 4 and 5 present the results and discussion.
For consistency with previous papers on this subject, 
we use a cosmology with $H_0$=50 km sec$^{-1}$ Mpc$^{-1}$,
$\Omega_m$=1.0, $\Omega_\Lambda$=0, throughout this paper, unless
stated otherwise.
 
\section{Young radio-loud AGN and the selection of the sample}

\subsection{Characteristics of  young radio-loud AGN}

In the course of the last decade, a class of extragalactic radio
source has been identified as being very young.
 This has opened many unique
opportunities for radio source evolution studies. Unfortunately,
 the nomenclature and use of acronyms in this 
field of research is rather confusing. This is mainly caused by
the different ways in which young radio-loud AGN are selected.
Selections of young sources are made in two ways, the first based  
on their broadband radio spectra, and the second based on their
compact morphology. A convex shaped spectrum, peaking at about 1 GHz 
distinguishes young radio sources from other classes of compact radio sources.
In this case they are called Gigahertz Peaked Spectrum ({\bf GPS})
radio
sources (eg. O'Dea et al. 1991, O'Dea 1998). Similar objects, which are
typically an order of magnitude larger in size, have their spectral turnovers 
shifted to the $10 - 100$ MHz regime, causing them to be dominated  at cm 
wavelengths by the optically thin parts of their spectra. 
These are called Compact Steep Spectrum ({\bf CSS}) radio
sources to distinguish them from the general population of extended 
steep spectrum sources (eg. Fanti et al. 1990). 

On the other hand, young radio sources are found in multi-frequency VLBI
surveys, in which they can be recognised by  compact jet/lobe-like
structures on both sides of their central core. They are called 
Compact Symmetric Objects ({\bf CSO}, Wilkinson et al. 1994). 
Their double sided structures
clearly distinguish them from the large majority of compact sources
showing one-sided core-jet morphologies.  Larger versions
of CSOs are subsequently called Medium Symmetric Objects ({\bf MSO}) and 
Large Symmetric Objects ({\bf LSO}).

The overlap between the classes of CSO and GPS galaxies is large and we 
believe that they can be considered to be identical objects.
However, a substantial fraction of GPS sources are optically 
identified with high redshift quasars, which in general show core-jet 
structures (Stanghellini et al. 1997). 
The relationship between GPS quasars and GPS/CSO galaxies
is not clear (Snellen et al. 1998b), and the former
may not be young objects et all. 

Although it was always speculated that GPS sources were young objects,
only recently has strong evidence been found to support this 
hypothesis. Monitoring several prototype GPS/CSO sources over a decade
or more using VLBI, allowed Owsianik \& Conway (1998) and Owsianik, Conway \& 
Polatidis (1998) to measure their hotspot advance speeds to be 
$\sim 0.1h^{-1}c$. These imply dynamical ages of typically $10^{2-3}$ years. 
Similar studies with similar results have now also been conducted by
Tschager et al. (2000) and Taylor et al. (2000).
Additional proof for youth comes from analysis of the overall radio
spectra of the somewhat larger CSS sources. Murgia et al (1999) show that 
their spectra can be fitted with synchrotron ageing models, implying 
ages of typically $10^{3-5}$ years.

The work of these authors indicate that GPS/CSO sources are very young and  
the likely progenitors of large, extended radio sources. However,
there may also be a population of young radio-loud AGN which are just
short-lived, and do not evolve into large size radio sources
(Readhead et al. 1994; Alexander 2000). In the latter case, the
connection of these host galaxies with those of other types of AGN 
would be less clear.

\subsection{Sample selection}

The sample of young radio sources used in this 
paper is based on the sample of bright GPS sources of 
Stanghellini et al. (1998). This complete sample of 33 
objects contains sources with turnover frequencies between 
0.4 and 6 GHz, with $\rm{S}_{\rm{5 \ GHz}} > 1$ Jy, and 
are located at $\delta > -25^{\circ}$ and galactic latitude 
$|b|>10^\circ$. Of this sample, 19 are optically identified 
with galaxies, of which 6 are found at $z<0.31$. Two objects, 
B1345+125 and B1404+286, were omitted from the sample, since
these were known to exhibit broad Seyfert-type emission lines 
in their optical spectra, implying a strong contribution 
of AGN related light. We further excluded B0428+205 due to 
its low galactic latitude and high foreground extinction.

To enlarge the usable sample and to be able to investigate 
possible correlations between host galaxy properties or 
optical AGN contribution and radio power, we included nearby
GPS galaxies from the sample of Snellen et al (1995) at
intermediate flux densities, and from Snellen et al. (1998a)
at faint flux densities. This involved the only GPS galaxy 
at low redshift from the intermediate sample, B1144+3517, 
and two nearby galaxies from the faint GPS sample,
B0830+5813 and B1946+7048. Details of the young radio-loud 
AGN are given in table \ref{details}.
Column 1 gives the B1950 name, and column 2 gives the J2000 position.
Column 3 gives the redshift. 
Column 4 gives the R band magnitude, taken from Snellen et al. (1996a),
O'Dea et al. (1996) and Snellen et al. (1998b). These are 
approximately corrected to the Cousins filter system, and 
galactic foreground extinction is taken into account, which is given 
in column 5 (Schlegel et al. 1998).
Their 5 GHz radio power, radio peak frequency, peak flux density,
angular and projected linear size are given in columns 6 to 10.

\begin{table*}
\begin{tabular}{ccccccclrr}
B1950 Name  & Position (J2000)   & z   &R&$E_R$ & $\rm{P}_{\rm{5 GHz}}$ & 
$\nu_{\rm{peak}}$ &  $\rm{S}_{\rm{peak}}$&$\theta$ & Size  \\
            &                    &     &mag&mag& Log W/Hz& GHz &Jy&mas&pc\\
\\
B0019$-$0001&00 22 25.5 +00 14 56&0.306&18.1&0.071&26.7&0.8&3.5&70& 390\\
B0830+5813  &08 34 11.1 +58 03 21&0.094&15.9&0.231&24.1&1.6&0.07&$<$5&$<12$\\   
B0941$-$0805&09 43 36.9 $-$08 19 31&0.227&17.5&0.076&26.4&0.5&3.4&50&231\\
B1144+3517  &11 47 22.1 +35 01 08&0.063&14.0&0.054&25.1&2.4&0.7&40& 66\\
B1819+6707  &18 19 44.4 +67 08 47&0.221&17.7&0.109&25.5&0.8&0.3&19& 86\\
B1946+7048  &19 45 53.5 +70 55 49&0.101&16.3&0.526&25.4&1.8&0.9&32& 80\\
B2352+4933  &23 55 09.4 +49 50 08&0.238&18.2&0.484&26.5&0.7&2.9&70&333\\
\end{tabular}
\caption{\label{details} Details of the young radio-loud AGN in the sample.}
\end{table*}

\section{Observations, data reduction and analysis}

\begin{figure*}
\psfig{figure=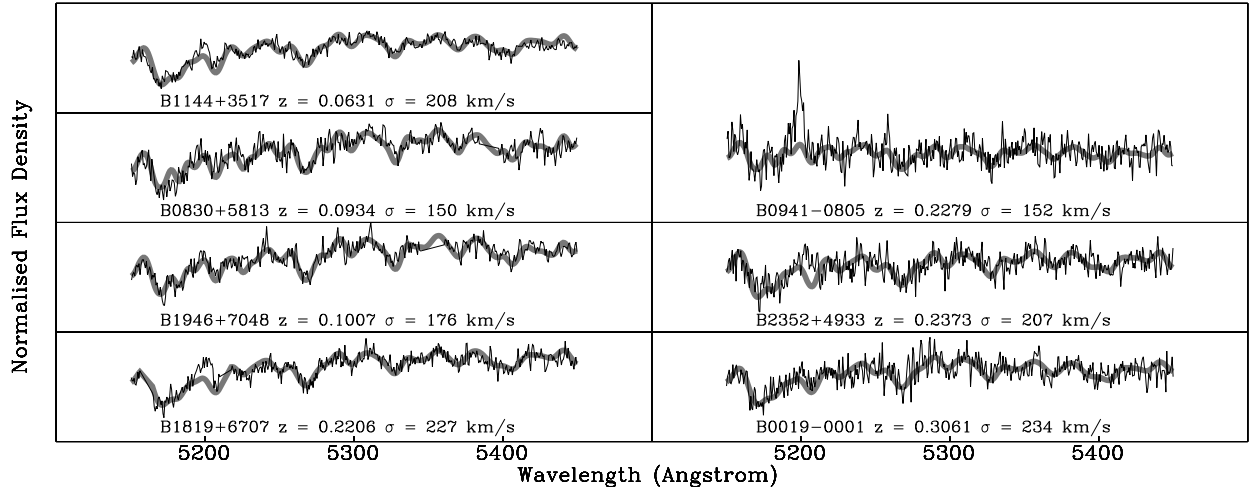,width=18cm}
\caption{\label{absorption} The part of the spectra used to 
measure the velocity dispersion,shown
 in order of increasing redshift. The thick
lines shows the best fit to the velocity broadened stellar spectra.
Note that in some cases, eg. for B0941-0805, the [NI]5199\AA$ $ emission line had
to be removed before the fitting procedure.}

\end{figure*}
\begin{figure*}
\psfig{figure=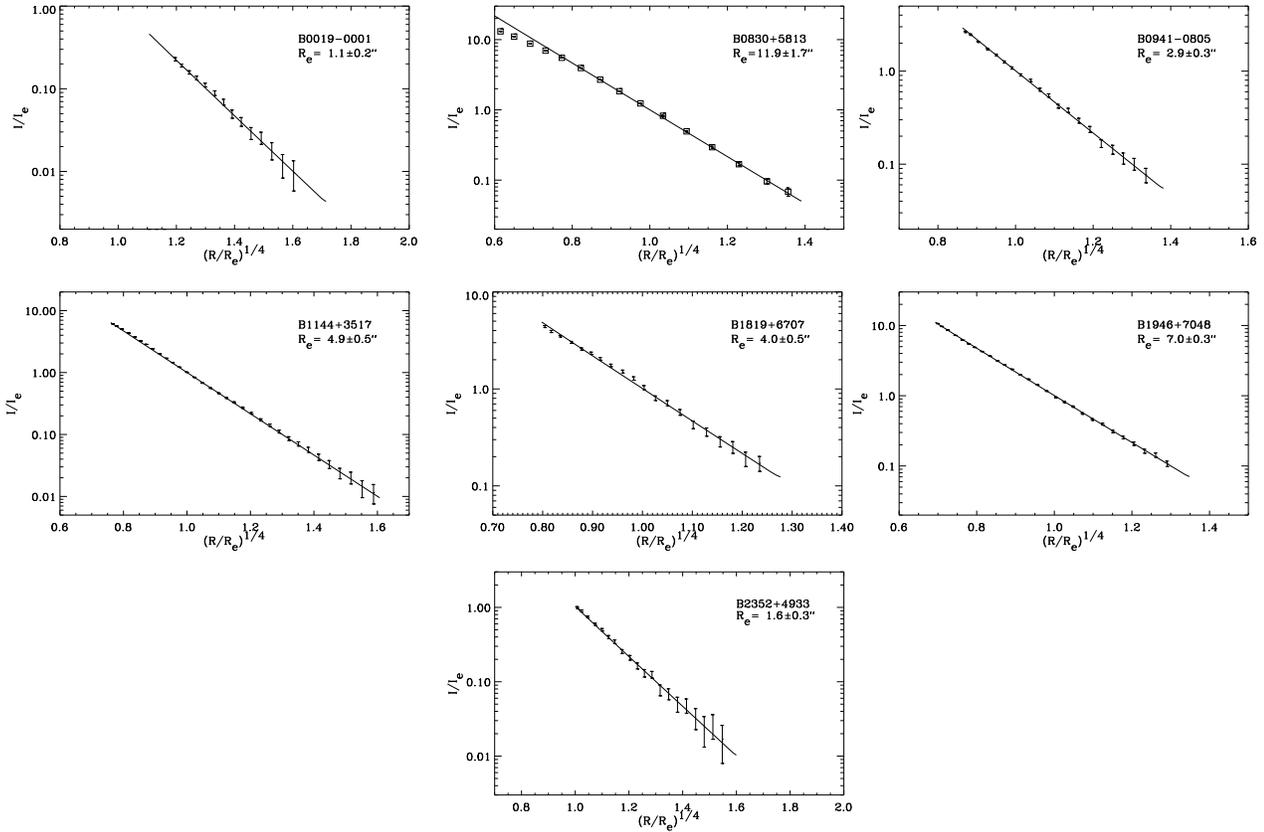,width=17cm}
\caption{\label{r4law} The normalised galaxy profiles with their best
fits. The profile of B0830+5813 is found to flatten in the central part.}
\end{figure*}
\begin{figure*}
\psfig{figure=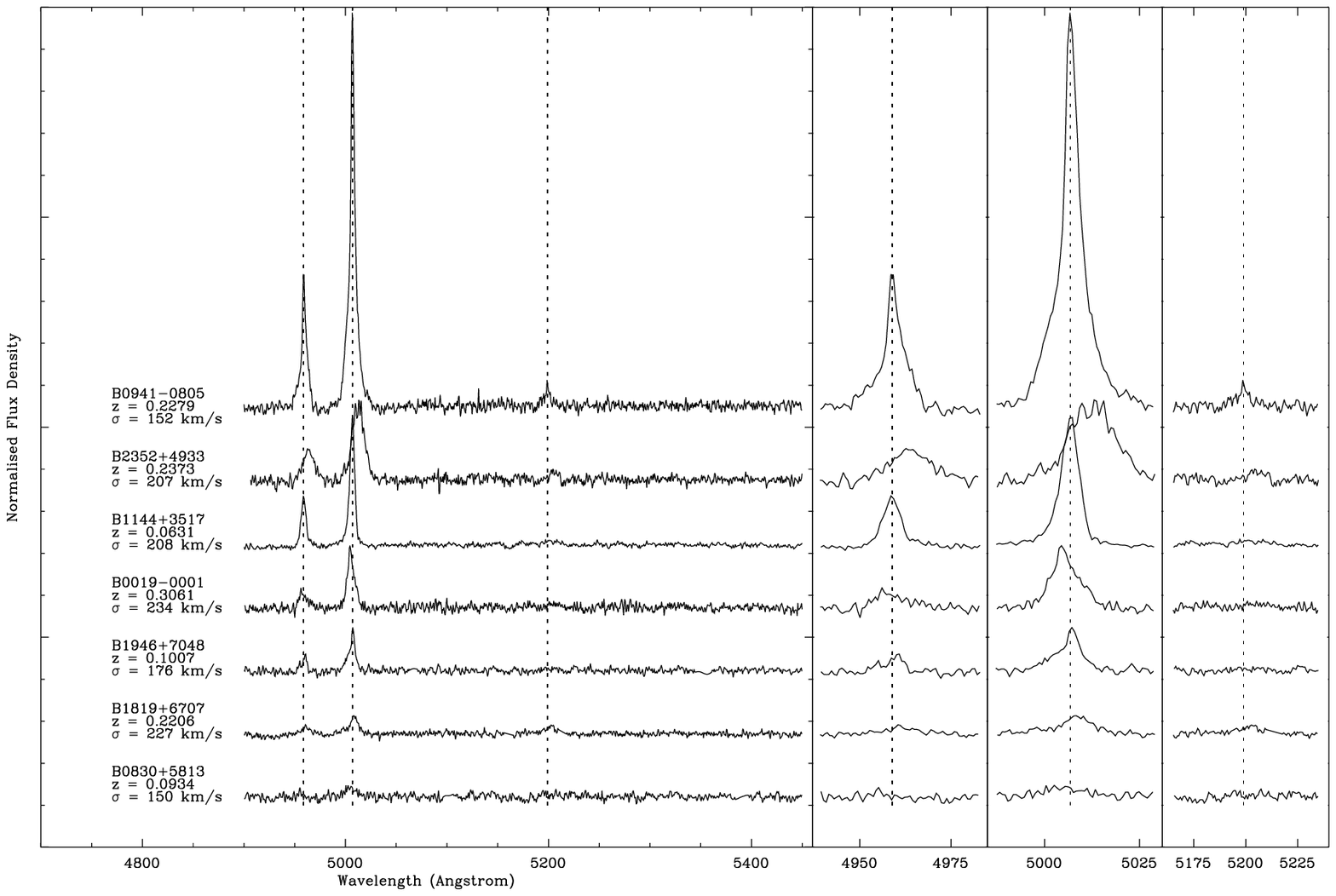,width=16cm}
\caption{\label{emission} The galaxy spectra of which the velocity
broadened
K-star spectra are subtracted. The three adjacent panels to 
the right zoom in on the [OIII]4959\AA, [OIII]5007\AA, and the 
[NI]5199\AA$\ $ emission lines, with the dashed lines indicating their 
rest-frame wavelengths, assuming the absorption line redshifts.}
\end{figure*}

The observations for this project are divided into 
three parts. 1) High-dispersion spectroscopy, to determine
the velocity dispersions and absorption line strengths of 
the objects. 2) Deep I-band CCD imaging to determine the 
surface profiles of the objects, and 3) B, V and R-band observations
to determine the rest-frame B-band effective surface densities of
these galaxies. 

\subsection{High-dispersion Spectroscopy}

The spectroscopic observations were carried out
using the 4.2-m William Herschel Telescope (WHT) located
at the Roque de Los Muchachos Observatory, La Palma in Spain.
Observations were carried out during two observing sessions
 July 29/30, 1997 and November 9-11, 1998. Both observing runs
were under non-photometric and variable seeing conditions.
Table \ref{log} gives the log of the observations, with in
column 1 the object name, and column 2 the observing 
run(s) during which the source was observed (1=July '97, 2=Nov '98). 
The exposure times per source ranged from 30 minutes to 
several hours, dependent on the redshift of the target and 
weather ( a significant proportion of the observing 
time was affected by clouds).

The red arm of the ISIS long-slit spectrograph was used,
with the R600R grating, which has 600 grooves/mm. 
The spectra were centred on the redshifted Mg\,b absorption feature
near 5170\AA.
The approximate central wavelengths are given in column 3
of table \ref{log}. The detector used was a Tektronic
$1024\times1024$ CCD, which resulted in a dispersion of 0.79 \AA$ $
per pixel, and a spatial resolution of 0.3 arcseconds.  
For all observations a slit-width of 
1.0 arcseconds was used. The spectral resolution for this setup 
was determined to be 1.6 \AA$ $ FWHM. The rest-frame 
spectral resolution for the observations of each target is 
given in column 4 of table \ref{log}.
The resulting signal-to-noise ratio per pixel is given 
in column 6.

\subsubsection*{Reduction and calibration of the spectra}

The data were reduced in IRAF.
The individual raw images were first bias subtracted.
After each target, an exposure was taken of the Tungsten 
lamp with the telescope pointing in the same position. 
This was normalised and  corrected for illumination effects,
and was used to flatfield the science images and to correct them 
for fringing. The data were then background subtracted and 
combined, and a one-dimensional spectrum retrieved.
Although the spectroscopic observations were not taken under photometric
conditions, observations
of spectroscopy standard stars were used to remove the response of the 
instrument from the spectra.

The wavelength calibration was performed 
using a CuAr calibration lamp. The wavelengths of the 
main sky-lines in the non-background subtracted images were
checked, and if necessary, a small shift in wavelength was
applied. Finally, the spectra were corrected to the 
heliocentric standard. The uncertainty in the 
absolute wavelength calibration is estimated to be 0.2\AA$ $ 
($\sim$10 km/sec).

\subsubsection*{The velocity dispersions and absorption-line 
redshifts}

The spectra were analysed by a Fourier Quotient method
(Simkin 1974; Sargent et al. 1977), using the software package
IDL. To determine the velocity dispersion and
the absorption line redshifts of the galaxies, their spectra
were compared to that of K-stars.
These stars proved to provide the most appropriate 
absorption line spectra for comparison to these old elliptical 
galaxies. For each galaxy, the continuum was subtracted
using a 5th order polynomial, after which the extremities of the 
spectrum were tapered using a Cosine-bell function.
The Fourier transform of this spectrum was divided by that of the 
continuum subtracted and tapered star spectrum. The result 
 represents the broadening function which was fitted by a Gaussian,
and gives the redshift, velocity dispersion, and relative
strength of the absorption lines of the galaxy to that of the 
template star. This fitting in Fourier space was typically performed
over a wave number range corresponding to 3-100\AA.
 Variations in redshift and velocity dispersion
between different template stars were found to be minimal.
In some cases the procedure was preceded by the removal of a  
small contribution from the [NI]5199\AA$ $ emission line.

Since elliptical galaxies exhibit radial gradients in their velocity
dispersions, the measured velocity dispersion had to be corrected
for the effective aperture size of the observations. We used the 
relation found by Jorgensen, Franx \& Kjaergaard (1995) to 
correct the velocity dispersions to an aperture diameter
of 1/4 of the effective radius, $r_e/4$.
This resulted in aperture corrections up to 20 km sec$^{-1}$. 
 
\subsubsection*{Mg\,b absorption line equivalent widths}

 \begin{table}
\caption{\label{log}Parameters of the WHT spectroscopic observations}
\begin{tabular}{ccccccclrr}
B1950 Name  & Obs.     & Cen.& $\sigma_{\rm{res}}$&SNR\\
            & run      & wave&                  &$pix^{-1}$\\
            &    & \AA & km/sec           &\\
\\		       
B0019$-$0001&1\&2&6800&30&15&\\
B0830+5813  & 2  &5650&36&19&\\   
B0941$-$0805& 2  &6350&32&13&\\
B1144+3517  & 1  &5500&37&35&\\
B1819+6707  & 1  &6350&32&26&\\
B1946+7048  &1\&2&5700&35&12&\\
B2352+4933  & 1  &6450&32&16&\\
\end{tabular}
\end{table}

The strength of the Mg\,b absorption feature was measured in 
an `index passband' flanked by two `pseudo-continua' 
(see Worthey 1994). 
The wavelength-ranges of the index and continuum band 
are as defined in Worthey et al. (1994, table 1).
Unfortunately, due to a varying contribution of the [NI]5199\AA$ $ 
emission line in the upper pseudo-continuum band of the Mg\,b index, 
the standard Mg\,b bands could not be
used. Instead, an upper pseudo-continuum band at 5205$-$5220\AA$ $ 
was used, avoiding this line. Our template star spectra
were used to calibrate this new measure of the Mg\,b index,
and we found the empirical relation
Mg\,b$=0.677+0.790\times$Mg\,b$_{\rm{alt}}$,
to correct these values.

The measured absorption indices had to be corrected for the velocity 
broadening of the spectra. This effect smears out the absorption lines,
possibly extending them to outside the index passband and 
into the pseudo-continuum bands, 
resulting in a  reduction of the measured absorption index. 
This effect, which is a function of the velocity dispersion,
is corrected for by estimating its influence by velocity
broadening the stellar-spectra. The corrections, in the order of 
10\%, were compared to that derived by Trager et al. (1998), and 
found to be similar.

Since elliptical galaxies have radial gradients in line strengths,
the effects of different aperture size must also be taken into 
account. We used a similar approach to that of Ziegler
\& Bender (1997), using Mg$_2$ profiles analysed by Gonzales 
\& Gorgas (1995). The values were corrected to a radius of 
$r_e$/4, resulting in an adjustment between
$-$0.25 and +0.28. 

\subsubsection*{The emission line equivalent widths and redshifts}

The best fitting velocity broadened stellar spectrum was subtracted 
from each galaxy to isolate the emission lines and measure
their equivalent widths. The velocity 
shifts of the emission lines were determined by comparing 
the measured peak wavelengths with that of laboratory
rest-frame measurements.

\begin{table*}
\caption{\label{results} The results of the imaging and spectroscopic observations.}
\begin{tabular}{cccrrccrr}
B1950 Name  & Redshift    &$\sigma_{e/4}$& \multicolumn{2}{c}{Effective Radius}  & $I_e$ &\multicolumn{2}{c}{Equivalent width}&Vel\\
            &             &    &      &      &                      & Mg b &[OIII]&shift  \\
            &             & km/sec& $''$ & kpc & mag & \AA &\AA  & km/sec\\
B0019$-$0001& 0.30615$\pm$0.00029&228$\pm$32& 1.1$\pm$0.2 &  6.4$\pm$0.8&21.85   &4.30& 17&-90\\
B0830+5813  & 0.09343$\pm$0.00012&146$\pm$17&11.9$\pm$1.7 & 28.0$\pm$0.8&25.11   &4.33&  3&-110\\
B0941$-$0805& 0.22790$\pm$0.00012&148$\pm$17& 2.9$\pm$0.3 & 14.0$\pm$0.9&24.03   &2.73& 88& +20\\
B1144+3517  & 0.06313$\pm$0.00013&202$\pm$18& 4.9$\pm$0.5 &  8.3$\pm$0.5&21.82   &4.11& 25& -5\\
B1819+6707  & 0.22071$\pm$0.00011&221$\pm$16& 4.0$\pm$0.5 & 18.8$\pm$1.5&22.85   &4.29&  6&+100\\
B1946+7048  & 0.10083$\pm$0.00009&171$\pm$20& 7.0$\pm$0.3 & 17.7$\pm$0.5&24.33   &4.64&  8& -0\\
B2352+4933  & 0.23790$\pm$0.00016&201$\pm$17& 1.6$\pm$0.3 &  7.9$\pm$1.0&22.15   &4.35& 35&+350\\
\end{tabular}
\end{table*}

\subsection{I-band CCD imaging}

The imaging observations to measure the 
effective radii of the galaxies were also carried out
using the WHT, this during the second observing session
at November 9-11, 1998. The weather was non-photometric
and the seeing conditions varied considerably. 
The auxiliary port camera at the Cassegrain focus was used
with a Tektronics 1024$\times$1024 CCD detector, which resulted
in a 0.11$''$ pixelsize and an unvignetted (circular) field 
with a diameter of 1.8'. After every exposure of a few minutes,
the telescope was offsetted by tens of arcseconds, to allow 
more accurate sky-subtraction. 
Total exposure times ranged from 10 to 45 minutes per galaxy. 

The data were reduced in IRAF.
The individual raw images were first bias subtracted and
flat fielded. These were then combined using median 
filtering to produce a fringe-images which was
then appropriately normalised and subtracted from the 
frames. The centroids of bright stars on the frames were used 
to determine their relative shifts, which were then combined
to produce the final image. No flux calibration was attempted
due to the non-photometric conditions.

\subsubsection*{The effective radii}

The reduced I band images were used to the determine the 
effective radii of the galaxies. This analysis was done
in the IDL software package. Firstly, the centroid of a 
galaxy was determined and its ellipticity fitted at an
isophote at 20\% of the value of the central peak. 
The level of intensity was measured for an 
increasing major axis, using the whole area of the ellipse,
in this way making full use of the data.
Some areas of the image, including nearby objects, were excluded
from the analysis. The central parts of the galaxies, which 
were influenced by the variable seeing, were also excluded.
The resulting profile was fitted by a $r^{1/4}$ 
de Vaucouleur profile.

\subsection{B,V and R-band CCD imaging}

Short CCD exposures of the objects were taken with
the Jacobus Kapteyn Telescope, which is also located 
at the the Roque de Los Muchachos Observatory, La Palma in Spain.
These served to determine their surface brightnesses at their 
effective radii in rest-frame B.
Observations were carried out during an observing session
from June 14 to June 23 2001, and during service observations
on July 4 and August 19 2001. All JKT observations were under 
good photometric conditions, and typical seeiing of 1$-$1.5$''$.
We used the CCD camera with the default SITe2 detector, which 
has 2048$\times$2048 24$\mu$m pixels, resulting in an image
scale of 0.33 arcsec/pixel.
We observed the objects in $B$ and $V$, or in 
$V$ and $R$ filter (depending on the redshift) for 2$\times$300 sec in
each filter. 
Standard star fields from Jorgensen (1994) were observed throughout
the nights at several zenith distances.

The data were reduced in a standard way, in a similar way as
the WHT I-band data. 
The photometric zero-points were found to vary by less then $<0.1$
magnitude between the different observing nights. 

\subsubsection*{Rest-frame B-band effective surface brightness}

To determine the effective surface brightness, $I_e$ in rest-frame B
band, first the effective surface brightnesses were measured in each
filter: the ellipticity and position angle of the galaxies as
determined from the deep I-band observations were used to determine
the surface brightness as function of radius for each galaxy,
excluding some areas containing neighbouring objects. This 1D profile
was then fitted with an $r^{1/4}$ law, using the effective radii
as found from the I-band observations, allowing $I_e$ and the
sky-level to vary. As a consistency check, the 
total intensities of the galaxies were also measured, using a large aperture
 of $>>r_e$. For a perfect 2-dimensional De Vaucouleur profile, these
relate as $I_{tot} = 22.66 I_e r_e^2$. For the galaxies 
free of nearby objects, they were found to match within 0.1 magnitude.

The $V$ and $R$ band flux density scales were corrected to match the $B$ band
zero-points, adopting an absolute spectral irradiance for
a zero magnitude object 
of $6.61\times10^{-9}$, $3.64\times10^{-9}$ and $2.25\times10^{-9}$ erg
cm$^{-2}$ sec$^{-1}$ \AA$^{-1}$ in the $B$, $V$ and $R$ band
respectively, and were also corrected for galactic foreground
extinction using the values of Schlegel et al. (1998). 
Interpolation between the surface brightnesses in the 
different filters was used to obtain the rest-frame B-band surface
brightness. This was then corrected for the cosmological dimming
and bandpass effects of $(1+z)^5$.

For two galaxies, B0830+5813, and B0914-0805, only R band
images were available. For these we applied a K-correction,
derived from the broadband colours of the other galaxies in the 
sample.

\section{Results and Discussion}

\begin{figure*}
\psfig{figure=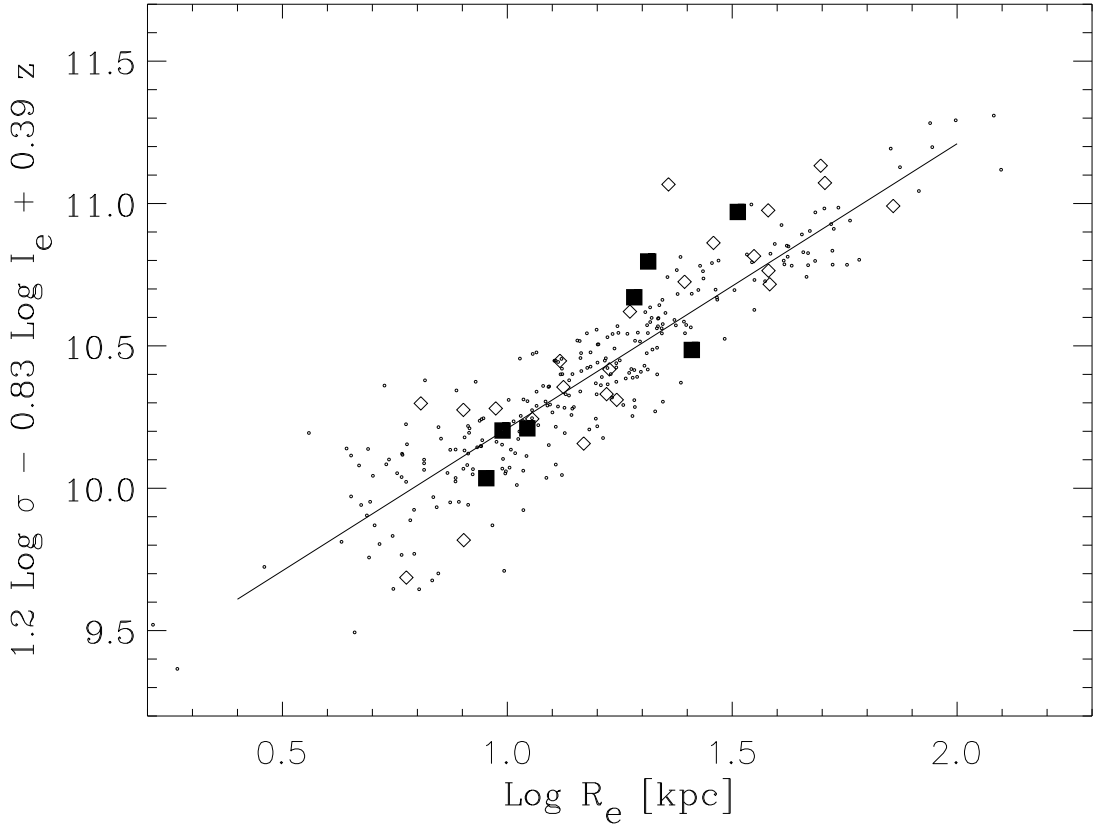,width=16cm}
\caption{\label{funplane} The fundamental plane of the host galaxies
of young radio-loud AGN (solid squares), with on the horizontal 
axis their effective radii, and on the vertical axis a combination 
of the velocity dispersion and their effective brightness, 
corrected for the cosmological evolution in the mass-to-light ratio 
as found by van Dokkum et al. (1998). The solid line indicates
the fundamental plane of elliptical galaxies, as derived from 
nearby galaxies by Faber et al. (1989; small circles).
The diamonds are the radio galaxies measured by Smith et al. (1990).}
\end{figure*}

The results of the imaging and spectroscopy are given in table 
\ref{results}.
Column 1 gives the B1950 name, columns 2 and 3 give the redshift and
the velocity dispersion corrected to an aperture of $r_e/4$. 
Column 4 and 5 give the effective radii in 
arcseconds and kpc. Column 6 gives
the brightness level at the effective radius for rest-frame B-band,
corrected for cosmological dimming and galactic foreground 
extinction.
Columns 7 and 8 give the rest-frame equivalent widths for the Mg b band (also corrected to an aperture of $r_e/4$)
and the [OIII]5007\AA\ emission line. 
Column 9 gives the velocity shift of
the [OIII] line with respect to absorption line redshift, where a
positive value indicates a redshift. The uncertainties in $I_e$ are
typically 0.05 magnitude, while the uncertainties in the equivalent
widths are $\sim10\%$. 

The part of the spectra used to measure the velocity dispersions of
the galaxies is shown in figure \ref{absorption}, in
order of increasing redshift. The thick lines indicate the best fit
to the velocity broadened K-star spectra. 
The measured velocity dispersions range from 
from 146 km sec$^{-1}$ for B0830+5813, to 228 km
sec$^{-1}$ for B0019-000.
Figure \ref{r4law} show the normalised galaxy profiles 
 with their best fits. The observed effective radii range
from 1.1$''$ for B009$-$0001, to 11.9$''$ for B0830+5813. Note that 
the profile of B0830+5813 slightly flattens in its inner part.
The best fitted velocity broadened K-star spectra were subtracted
from those of the galaxies to obtain their emission line spectra. 
These are shown in figure \ref{emission}. 
The three adjacent panels to the right zoom in on the 
[OIII]4959\AA, [OIII]5007\AA, and the [NI]5199\AA$ $ emission lines,
 with the dashed lines indicating their 
rest-frame wavelengths, assuming the absorption line redshifts. 
it shows that the emission line redshifts correspond well with those
infered from the absorption lines, except for B2352+4933, which 
emission lines are clearly redshifted by 350 km/sec.

\subsection{The Fundamental Plane}

\begin{figure}
\psfig{figure=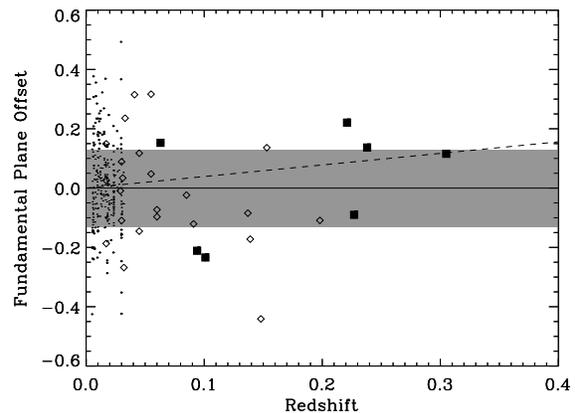,width=8cm}
\caption{\label{funevol}The offset of the host galaxies 
to the local fundamental plane for cluster ellipticals as function of
redshift. The dotted line is the evolution in mass-to-light ratio as
found by van Dokkum et al. (1998). The symbols are as in figure \ref{funplane}. 
There is a possible hint that the more distant galaxies are 
brighter, but more data are needed to confirm this.}
\end{figure}
\begin{figure}
\psfig{figure=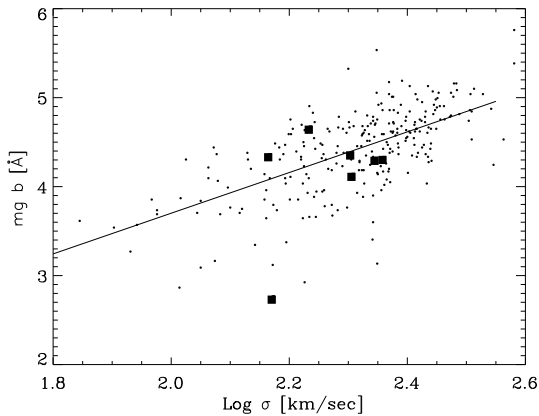,width=8cm}
\caption{\label{sigmaMgb}The Mg\,b-$\sigma$ relation for the host
galaxies of young radio-loud AGN (squares). 
The solid line indicates the Mg\,b-$\sigma$ relation for 
local ellipticals from Faber et al. (1989; small circles).}
\end{figure}

\begin{figure}
\psfig{figure=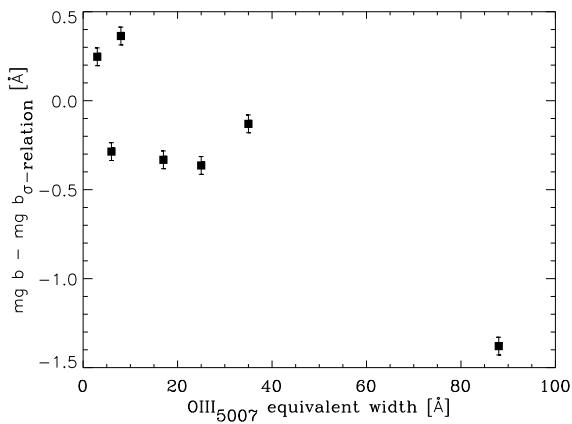,width=8cm}
\caption{\label{OIII_Mgb}The offsets of the Mg\,b indices to 
the Mg\,b-$\sigma$ relation for local ellipticals by 
Faber et al. (1989), as function of [OIII] equivalent width. 
The correlation indicates that the Mg\,b index has likely 
been influenced by continuum contamination from AGN related
light in the nuclear spectra.}
\end{figure}

The main aim of the project was to investigate whether young 
radio-loud AGN could be used to study the cosmological evolution 
of the fundamental plan for ellipticals, being 
selected independently of optical selection biases.
Figure \ref{funplane} shows the fundamental plane with the angle of 
projection in the 
$\sigma$, $I_e$, $r_e$ - space chosen to be similar to that used
in previous studies, with on the horizontal axis the logarithm of 
the effective radius, $r_e$ (in kpc), and on the vertical axis 1.2 
times the logarithm of the velocity dispersion $\sigma$ 
(in km sec$^{-1}$) minus 0.83 times the logarithm of the brightness
level at the effective radius, $I_e$, with $I_e=10^{-0.4\mu_e}$.
The data are corrected for the cosmological evolution in 
mass-to-light ratio as found by van Dokkum et al. (1998), 
corresponding to $\Delta$mag = 1.22$\times z$.
The solid squares show the galaxies in the sample, and
the diamonds are the radio galaxies from Smith, Heckman \& 
Illingworth (1990). The small circles are the local ellipticals
from Faber et al. (1989) with heliocentric velocities $>$1500 km/sec,
with the solid line their best linear fit. The galaxies in our 
sample follow the fundamental plane well, with an average offset
of 0.01 and a dispersion of 0.17, similar to that of the Smith
et al. sample, but with a slightly higher dispersion than the local
galaxies in the Faber et al. sample (0.13).

To investigate whether evolution in M/L is visible over the small
redshift range in our sample, the offsets of the datapoints to
the local fundamental plane, $\Delta$FP, as function of redshift 
is shown in figure \ref{funevol}. The symbols are as in figure \ref{funplane},
with the grey band indicating the dispersion in the local fundamental 
plane, and the dashed line showing the evolution as found by 
van Dokkum et al. (1998) for cluster ellipticals.
The galaxies containing young radio sources located at z$>$0.2 seem to
have on average a slightly positive offset from the local fundamental
plane, with $\Delta$FP=0.10$\pm$0.06. Therefore
evolution is only detected at a 1.7$\sigma$ level in this small dataset.
Clearly, more data are needed, preferably out to higher redshift,
to confirm evolution, and to test the findings by van Dokkum et al.

\subsection{The Mg\,b - $\sigma$ relation}

The absorption line strengths of a galaxy also give a good,
independent insight in its stellar population.
The Mg\,b-$\sigma$ relation, and its dependence on redshift,
has been shown to be a powerful tool to study galaxy evolution
(Ziegler \& Bender 1997).
The Mg\,b-$\sigma$ relation for the galaxies in our sample is
shown in figure \ref{sigmaMgb}. The small circles are the data
from the local Faber et al. sample, where the best fit
is indicated by the solid line. 
These datapoints were converted Mg$_2$ magnitudes, using
the relation Mg\,b\AA$^{-1}$=15.0Mg$_2$ (Ziegler \& Bender 1997). 
The Mg\,b index of B0941-0805 is clearly lower
than expected from the local relation. Since this object exhibit
a strong emission line spectrum, we suspect that this Mg\,b deficit
is caused by a contamination of the continuum by AGN related light.
In figure \ref{OIII_Mgb}, the offsets of the Mg\,b indices to
the local Mg\,b-$\sigma$ relation is shown to be clearly correlated
with OIII$_{5007}$ equivalent width, indicating that AGN 
contamination indeed plays a role. The Mg\,b deficit in B0941-0805
can be explained by an AGN continuum contribution of 35\%.
For the other galaxies in the sample the AGN contribution is 
less than $<$5-10\%. Note that this AGN contamination is for 
nuclear spectra, using a narrow slit. It is unlikely to have 
any effect on the determinations of the surface brightness 
profiles as done at and beyond $r_e$, and it is therefore improbable
to affect the determination of the fundamental plane parameters. 
However, this AGN contribution, although generally small, 
makes investigating galaxy evolution using the Mg\,b-$\sigma$ relation not 
very useful for radio galaxies. 

\subsection{Black hole masses and their connection with radio power}

\begin{figure*}
\psfig{figure=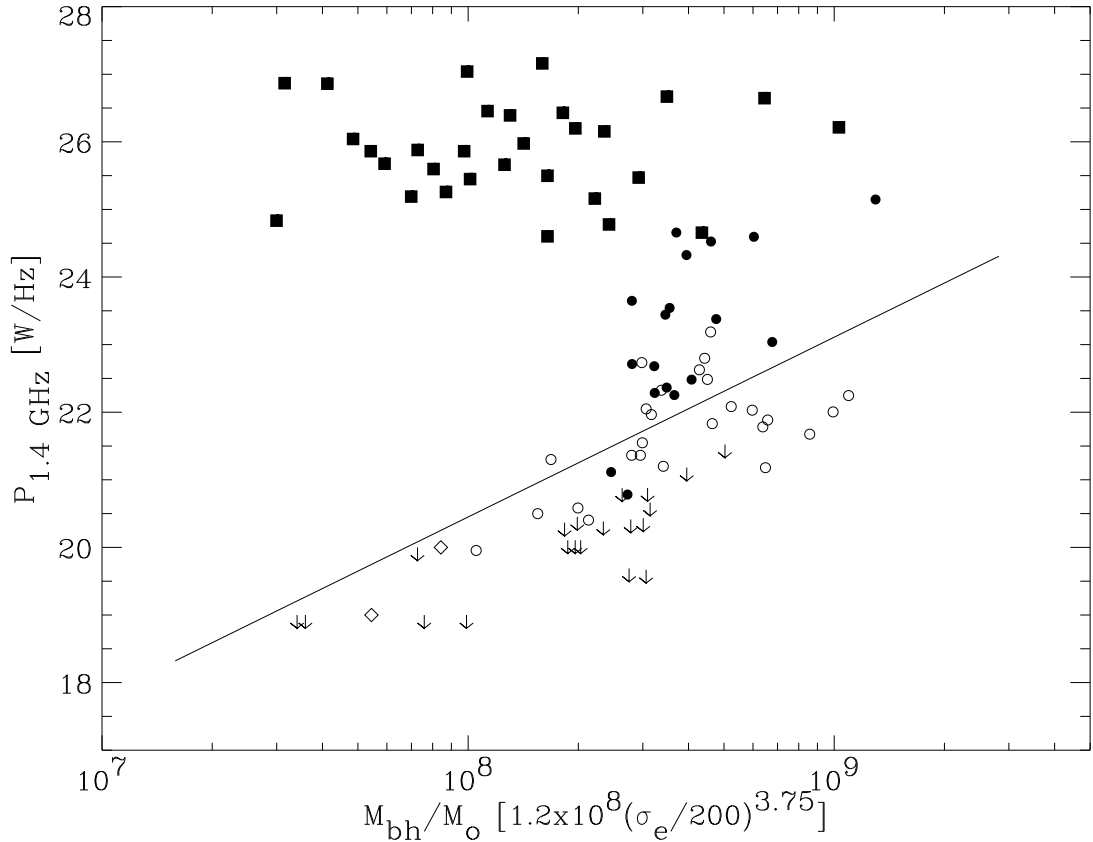,width=16cm}
\caption{\label{Mbhradio} The radio luminosity at 1.4 GHz versus 
black hole mass, as derived from the central velocity dispersion. The 
squares are those objects selected in the radio 
(from this paper and Smith et al. 1990). The open and closed circles are 
optically selected galaxies from the Faber et al. (1989) sample, with the 
former being point sources in the NVSS, and assumed to be radio-quiet, and 
the the latter being extended in the NVSS and assumed to be radio-loud AGN.
The few diamonds at low black hole mass are galaxies added from the local 
group. The solid line is the $M_{bh}-P_{\rm{radio}}$ relation as found by 
Franceschini et al. (1998).}
\end{figure*}

The work by Gebhardt et al. (2000) and Ferrarese \& Merrit (2000)
have shown that the mass of a central black hole is strongly
correlated with the central velocity dispersion of its host galaxy. 
According to Gebhardt et al. the relation follows
\[ M_{\rm{bh}}=1.2\times10^8\times \left(\frac{\sigma_e}{200 \ 
\rm{km/s}}\right)^{3.75} M_\odot\]
which is found to have a lower dispersion than the correlation with  
absolute (bulge) magnitude. This $M_{bh}-\sigma$ relation allows us to 
infer black
hole masses from the central velocity dispersion measurements, 
for the galaxies in our sample, and those in Smith et 
al. and Faber et al. 
Uncertainties in the black holes masses are dominated by the scatter
in the $M_{bh}-\sigma$ relation, which is estimated to be a factor 
of $\sim$2 (Gebhardt et al. 2000).

A variety of authors have discussed the relation between $M_{bh}$ 
and radio power. Long before the correlations between 
black hole mass and central velocity dispersion or galactic bulge mass 
were established, evidence has been provided of a relation
between galactic bulge luminosity and/or velocity dispersion 
with radio luminosity $L_R$ (Heckman 1983, Nelson \& Whittle 1996). 
Franceschini et al. (1998) showed that for a small sample of 
nearby galaxies, the non-thermal radio emission is strongly 
correlated with the black hole mass, from the dwarf elliptical M32 to 
the powerful radio galaxy M87, with a surprisingly small scatter,
which was confirmed by McLure et al. (1999).
More recently, other studies (eg. Laor 2000; Ho 2002) show that AGN are in 
general located above the overall $M_{bh}-L_{R}$ relation as found by 
Franceschini et al. Interestingly, Dunlop and McLure and collaborators
(Dunlop et al. 2002; Dunlop \& McLure 2002) have suggested that there is 
an $M_{bh}$-dependent lower and upper limit to the possible radio output of a 
black hole, offsetted from each other over several orders of magnitude.

To investigate these issues further, and in particular the relative position
of radio galaxies on the $M_{bh}-L_{R}$ relation, we correlated
the galaxies from Faber et al. with the NVSS-VLA 1.4 GHz survey 
(Condon et al. 1998) and determined their radio luminosities and sizes.

The $M_{bh}-L_{R}$ relation for the galaxies in the Faber et al. 
(circles), and the radio selected galaxies from Smith et al. and this paper
(squares) are shown in figure \ref{Mbhradio}. The galaxies 
in Faber et al. which exhibit extended radio structure, and therefore
clearly identified as active galaxies, are indicated by solid 
circles. A few objects from the local group are added, and are indicated
by diamonds.

To avoid the diagram being crowded by meaningless upperlimits,
only those objects are plotted, if their radio flux would have been
detectable in the NVSS if they radiate at a luminosity expected
from the Franceschini relation.
Although this produces a varying redshift cut-off with 
black hole mass, it does not bias the result towards particular 
radio luminosities. Clearly, the radio luminosities of the 
presumably inactive galaxies correlate well with black hole mass,
although they seem to be slightly fainter than expected
from the Franceschini relation, and have a larger scatter, with
the radio power of  inactive galaxies at a given black hole mass 
ranging more than 2 orders of magnitude. 
The active galaxies (those selected in the radio, and those 
selected from the Faber et al. sample exhibiting extended NVSS emission), 
do not show any sign of a correlation with black hole mass. At a given
$M_{bh}$ the radio luminosity can span over 7 orders of magnitude,
making radio power unsuitable as a black hole mass estimator. 
We note that the black hole masses for radio galaxies as 
derived by Dunlop et al. (2002) are on average about a factor of 4 higher
than given here (assuming $M_{bh}$=0.0013$M_{\rm{bulge}}$). 
This is caused by their different method used to determine $M_{bh}$, based
on the bulge luminosity $-$ bulge mass $-$ black hole mass relation:
the average absolute R-band magnitude for the Smith et al. sample 
($-23.71\pm0.12$, assuming V-R=0.6) is the same as that for the 
radio galaxies in the Dunlop et al. sample ($-23.66\pm0.16$), indicating that 
the properties of the host galaxies in the two samples are the same. 
Note that the differences between these methods result mainly in 
an uncertainty in the absolute values of $M_{bh}$,
not in their relative values. All the radio galaxies are within the 
allowed range of radio power for their black hole masses as proposed 
by Dunlop et al. (2002).

\subsection{Connecting the local populations of radio galaxies and inactive 
ellipticals}

Since the radio galaxies are selected virtually independently of their 
optical properties, and the optically selected galaxies independently of 
their radio properties, the samples can be used to connect the 
two populations.
Figure \ref{Mbhradio} shows that there is a clear difference between the 
$M_{bh}$ distribution for radio selected
and optically selected AGN, with those selected in the radio, 
being biased towards lower black hole masses. We believe this is the 
natural consequence of the Faber et al. sample strongly favouring 
intrinsically more luminous galaxies, caused by its apparent magnitude
limit. In contrast, the selection of objects on radio 
emission is practically volume based, with the parent population of 
potential radio-loud AGN containing relatively many more galaxies with
lower velocity dispersions and black hole masses than the Faber et al.
sample, which causes the sample of radio-selected AGNs to cover
more galaxies with low $M_{bh}$.

\begin{figure}
\psfig{figure=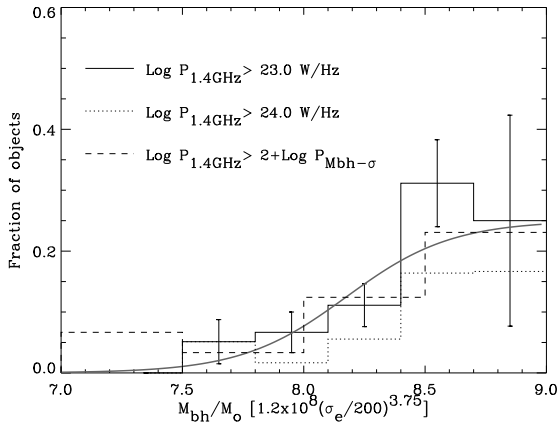,width=8cm}
\caption{\label{fraction} The fraction of radio-loud AGN as function of
black hole mass, as derived from the optically selected sample of Faber 
et al. (1989). The distributions are explained in the text. The smooth 
curve is the function P(AGN|$M_{bh}$) used to determine the density
and $M_{bh}$ distribution for radio-loud AGN with $P_{\rm{1.4GHz} > 23.0}$
W/Hz.}
\end{figure}

\begin{figure}
\psfig{figure=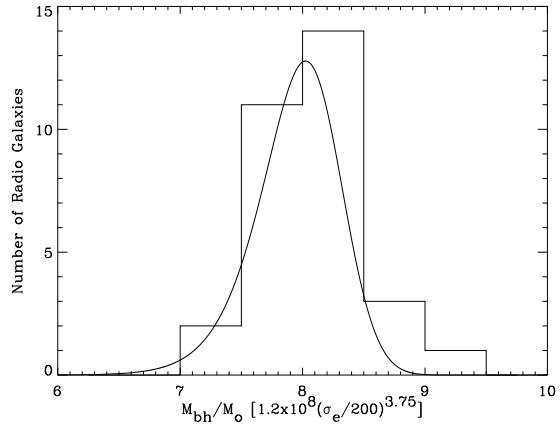,width=8cm}
\caption{\label{bh_dist}The distribution of black hole masses for the 
radio selected galaxies. The smooth curve (arbitrarily scaled) shows the 
predicted distribution of $M_{bh}$ derived from the local elliptical 
galaxy luminosity function
and the fraction of black holes with powerful radio sources 
as function of $M_{bh}$ as derived from the optically selected sample.}
\end{figure}

These black hole mass distributions form the key to understand the 
connection between the populations of radio-loud and radio-quiet galaxies.
First we determined the fraction of radio-loud AGN in the Faber et al. sample 
as a function of black hole mass. This is shown in figure \ref{fraction}, with 
the solid and dotted lines indicating the fraction of galaxies with 
$P_{\rm{1.4GHz}}$ $>23.0$
and $>24.0$ W/Hz respectively, the former decreasing from about 1/5 at
$M_{bh}=10^9M_\odot$ to $<1/20$ at $M_{bh}=10^{7.5}$. Note that 
a different subsample has been used for this than in figure \ref{Mbhradio},
with such a distance limit that it can be determined from the NVSS whether
 the galaxy has a radio power above or below the cut-off.
The dashed line shows the fraction of objects as function of $M_{bh}$ with 
a radio power more than 100 times that expected for the 
$M_{bh}-P_{\rm{radio}}$
relation by Franceschini et al (1998). It shows that although the 
radio selected AGN do not have black holes with M$_{bh}<3\times10^7M_{\odot}$,
black holes below this mass can still be active but only with a very low 
probability. Several objects with $M_{bh}\sim10^7M_\odot$  have been found 
with extended double-lobed low power radio sources in the Faber et al. 
sample.

With this estimate of the fraction of black holes producing powerful 
radio sources as a function of $M_{bh}$, the populations of radio-loud 
and radio-quiet galaxies can be linked. The absolute magnitude of a 
galaxy is connected to its central velocity dispersion by the Faber-Jackson
relation (Faber \& Jackson 1976), which in itself is related to the 
black hole mass via the $M_{bh}-\sigma$ relation. In this way, the local
distribution of black hole masses can be roughly deduced from the local
galaxy luminosity function. The combination of the Faber-Jackson relation, 
which was re-determined for $\sigma_e$ from the galaxies in the Faber et al. 
sample, and the $M_{bh}-\sigma_e$ relation, results in a local black hole 
mass distribution of 
\[
\phi(M_{bh})dM_{bh} =0.65 \frac{\phi^\star}{M^\star_{bh}}\rm{Exp} 
 \left( -\frac{ M_{bh} }{M^\star_{bh}} \right)^{0.65}\left(\frac{M_{bh}}{M^\star_{bh}}\right)^{0.65(1+\alpha)-1} \hspace{-1.2cm} dM_{bh}
\]
with $\rm{Log} \ M^{\star}=-(8.818+M_B^\star)/1.616$, where $M^\star_B$, $\alpha$, and $\phi^\star$
are the Schechter Function parameters of the local luminosity function in
B band. 
Our estimate of the fraction of black holes
producing powerful radio sources should now link this distribution of $M_{bh}$ to
the local {\it radio} luminosity function.
For our calculation we assume that the fraction of radio-loud black
holes with $P_{\rm{1.4GHz}}>23.0$ W/Hz is $0.25\times(1+(M_{bh}/1.5\times10^8M_\odot)^2)$, which is 
represented by the smooth curve in figure \ref{fraction}.
Using the Schechter parameters for the local luminosity function
of {\it elliptical} galaxies only, as derived from the 2dF Galaxy Redshift Survey
(type 1 gals; Madgwick et al., 2002) we derive a local radio source
density of $1.6\times10^{-5}$ Mpc$^{-3}$ at Log P$_{\rm{1.4GHz}}>23.0$ W/Hz.
This corresponds very well to the density of radio sources above this 
luminosity cutoff as derived from the local radio luminosity function, 
which is $1.3\times10^{-5}$ Mpc$^{-3}$ (eg. Dunlop \& Peacock, 1990).
In addition, the distribution of the black hole masses in the radio selected
 samples is shown in Figure \ref{bh_dist}. Although these samples are not 
complete, we believe that they are roughly selected at random from the 
general population of 
radio sources with Log P$_{\rm{1.4GHz}}>24.0$ W/Hz. The smooth curve
represents the expected distribution of black hole masses from the 
calculations above, which is strikingly similar. These results confirm
that elliptical galaxies must comprise the large majority of the radio-loud 
population of active galaxies.

\section{Summary}

We have determined the fundamental galaxy parameters for a 
sample of 7 host galaxies of young radio-loud AGN at z$<$0.3, and
have shown that they follow the fundamental plane of elliptical
galaxies. There is some indication that the more distant objects
in the sample have higher M/L ratios, as expected from stellar 
evolution. As these galaxies with young radio sources are relatively
easy to select, and the contributions of the AGN light to the 
optical spectra are minimal, they can readily be used to study
the evolution of the fundamental plane out to z=1, independent
of optical selection effects.

Black hole masses are determined for the young radio-loud AGN,
and for samples of classical radio galaxies and optically 
selected ellipticals, using the $M_{bh}-\sigma$ relation.
We confirm the relation between radio power and black hole mass
as found by Franceschini et al. (1998),
but only for relatively passive ellipticals. Active galaxies do not 
show such a correlation and can range over 7 orders of magnitude in 
radio power at a given $M_{bh}$.

The local galaxy luminosity function was combined with the 
Faber-Jackson relation and the $M_{bh}-\sigma$ relation to obtain
the local distribution of black hole masses. This, combined with 
the fraction of black holes with powerful radio sources as function of 
$M_{bh}$, which was determined from the optically selected sample,
results in a radio source density and radio-loud black hole mass
distribution. These are found to be consistent with the local radio 
luminosity function and the distribution of black hole masses in the 
radio-selected samples, and confirms that elliptical galaxies must comprise 
the large majority of the radio-loud population of active galaxies.

\section*{Acknowledgements}
The William Herschel Telescope, the Isaac Newton Telescope, and
the Jacobus Kapteyn Telescope are operated on the island of La Palma by 
the Isaac Newton Group in the Spanish Observatorio del Roque de los
Muchachos of the Instituto de Astrof\'{i}sica de Canarias.
We wish to thank the service team of the ING and Dr. Karl-Heinz Mack for 
taking the JKT data.

{}
\end{document}